\documentclass{article}
\begin{document}
\title{ Pointlessness and dangerousness of the postulates of quantum mechanics.
\author{Jacques Moret-Bailly \footnote{Laboratoire de physique, Universit\'e de Bourgogne, BP 47870, F-
21078 Dijon 
cedex, France. jmb@jupiter.u-bourgogne.fr}}}
\maketitle

{\it Abstract

The formalism of quantum mechanics produces spectacular results, but its rules, its parameters are empirical, 
either deduced from classical physics, or from experimental results rather than from the postulates. Thus, 
quantum mechanics is purely phenomenological; for instance, the computation of the eigenvalues of the 
energy is generally a simple interpolation in the discrete space of the quantum numbers. The attempts to 
show that quantum electrodynamics is more precise than classical electrodynamics are based on wrong 
computations. The lack of paradoxes in the classical theory, the appearance of classical, true 
interpretations of the wave-particle duality justify the criticism of Ehrenfest and Einstein.

The obscurity of the quantum concepts leads to wrong conclusions that handicap the development of 
physics. Just as building a laser was considered absurd before the first maser worked, the concept of 
photon leads to deny a type of coherent Raman scattering necessary to understand some redshifts of 
spectra in astrophysics, and able to destroy the two fundamental proofs to the expansion of the universe.
\medbreak

R\'esum\'e

Le formalisme de la m\'ecanique quantique conduit \`a des r\'esultats spectaculaires, mais ses r\`egles et ses 
param\`etres sont empiriques, d\'eduits de la physique classique ou des r\'esultats exp\'erimentaux plut\^ot 
que des postulats. Ainsi, la m\'ecanique quantique est une pure ph\'enom\'enologie, par exemple le calcul 
des valeurs propres de l'\'energie est une interpolation sur l'espace discret des nombres quantiques. Les 
efforts tendant \`a d\'emontrer que l'\'electrodynamique quantique est plus pr\'ecise que 
l'\'electrodynamique classique d\'emontrent seulement qu'il convient d'\^etre rigoureux. L'absence de 
paradoxes en th\'eorie classique, l'apparition d'interpr\'etations classiques v\'eritables de la dualit\'e onde-
corpuscule paraissent une justification des critiques d' Ehrenfest et Einstein.

L'obscurit\'e des notions quantiques am\`ene \`a des conclusions inexactes qui handicapent le 
d\'eveloppement de la physique; ainsi, de m\^eme que le concept de laser a \'et\'e ni\'e avant la 
d\'emonstration exp\'erimentale, la notion de photon conduit \`a nier une forme d'effet Raman coh\'erent 
n\'ecessaire \`a l'explication de rougissements de spectres en astrophysique et susceptible de d\'etruire les 
deux preuves fondamentales de l'expansion de l'univers. 

}\bigbreak

{\leftskip=4cm 

"Llewlyn Thomas, a noted Columbia theorist, told me flatly that the maser could not, owing the basic physics 
principles, provide a pure frequency[ ]; so certain was he that he more or less refused to listen to my 
explanations[ ].On visiting Niels Bohr, the pioneer of quantum mechanics, in Denmark, he exclaimed: "But 
that is not possible"[ ] John von Neumann declared "That can't be right"[ ]. To physicists steeped in the 
uncertainty principle, the maser's performance made no sense at all" (C. H. Townes \cite{Townes}).
\par}\medskip
\section{Introduction.}
At the beginning, relativity and quantum mechanics were strongly criticised; but, while relativity is now 
widely accepted, so many physicists share now the scepticism of Ehrenfest and Einstein about quantum 
mechanics that the supporters of this theory look for an absolute proof of its necessity. On the contrary, 
this paper tries to show that the postulates of quantum mechanics must be rejected.

Quantum mechanics was an important tool in the development of physics in the twentieth century, but using 
the {\it formalism} of quantum mechanics does not require the {\it postulates}.

The following section shows that the symmetry properties often considered as a part of the formalism of 
quantum mechanics may be justified classically, while the computation of the eigenvalues of the energy 
uses so disparate methods that it appears phenomenological, and may be bound to either theory. It is only 
examples because they are too many applications of quantum mechanics.

The next section reinforces the previous one in the field of optics: the defenders of quantum mechanics 
performed experiments to prove an inadequacy of the classical theory but they used naive hypothesis, so 
that their demonstrations fail. The flimsiness of quantum mechanics leads to introduce absurd concepts, 
such as the photon, as a particle, leading to an absurd wording of the EPR experiment. At low lighting, the 
linearisation of optical effects in function of the amplitude of the electric field appears as a nearly trivial but 
very useful property.

Next section is a tentative explanation of the wave particle duality by (3+0)D solitons whose existence is 
demonstrated.

It appears finally that the photon leads the astrophysicists to reject the coherent Raman scattering (just as the 
maser was rejected), giving them the two main (and probably fallacious) proofs of the expansion of the 
universe. Consequently, they need to imagine fantastic explanations about observations on quasars and 
other massive objects.
\section{Classical justification of the formalism of quantum mechanics.}
A criticism of quantum mechanics is difficult because the frontiers of this theory are not well defined. The 
symmetries are often introduced by ''active transformations'' in which the particles are moved by the 
physicist; the active transformations are, as often claimed, a concept of quantum mechanics. In classical 
physics, the theoretician observes, does not change the system he observes, using ''passive 
transformations'' which act on mathematical tools, such as reference frames, only.
\subsection{Molecular symmetries}
The molecular symmetries are an easy to explain example of symmetries \cite{Morets} generally studied by 
''active transformations''. Show that classical physics introduces naturally the rules postulated by the 
theory of active transformations.

The classical problem is setting, at a given instant, the variables which allow to describe a molecule made of 
punctual atoms, taking into account the hypothesis of the existence of a remarkable configuration (called 
here equilibrium configuration), generally a configuration for which the potential function is minimal.

Suppose first that the molecule is made of different atoms $A, B, C,...$ of masses $m_A, m_B, m_C, \dots$. 
The equilibrium configuration is a geometrical, solid figure of points $a, b, c,...$, defined within a 
displacement. The real molecule is distorted; the equilibrium configuration must be bound to the molecule, 
for instance to study small movements, considering that the atoms $A, B, C,...$ are displaced from the 
corresponding points $a, b, c,...$. For instance, this binding is done using the conditions set by Eckart: 
First a centre of mass $O$ of the equilibrium configuration is defined by
\begin{equation}
m_a \overrightarrow{Oa}+ m_b \overrightarrow{Ob}+ m_c \overrightarrow{Oc}+..=\overrightarrow{0},
\end{equation}
 The first condition sets that $O$ is the centre of mass of the molecule:
\begin{equation}
m_a \overrightarrow{aA}+ m_b \overrightarrow{bB}+ m_c \overrightarrow{cC}+..=\overrightarrow{0},
\end{equation}
The second condition is:
\begin{equation}\overrightarrow{Oa}\wedge (m_a \overrightarrow{aA})+\overrightarrow{Ob}\wedge ( m_b 
\overrightarrow{bB})+ \overrightarrow{Oc}\wedge (m_c \overrightarrow{cC})+..=\overrightarrow{0}.
\end{equation}
A mobile reference frame $O'xyz$ may be bound to an equilibrium configuration independent of the molecule, 
obtaining a ''reference configuration" which is a mobile, geometrical solid; the reference configuration may 
be defined by a table of coordinates which will be the components of $\overrightarrow{O'a}, 
\overrightarrow{O'b}, \overrightarrow{O'c},\dots      $ in $O'xyz$. The frame is bound to the molecule 
superposing by a displacement the reference configuration with the equilibrium configuration bound to 
the molecule by the Eckart conditions. The relative coordinates are defined.

If, in the set $A, B, C,...$ two or more points have the same mass (with an assumed approximation), they will be 
designed by the same letter, so that the displacement vectors such as $\overrightarrow{aA}$ are not 
uniquely defined; at a given time, the ambiguity may generally be solved setting, for instance that the sum 
of the modulus of the displacement vectors is minimal. If, during the movement, the definition of the 
displacement vectors is not changed, the molecule is said ''semi-rigid''. To define the coordinates, indices 
distinguish the $a, b, c \dots$ points (thus the displacement vectors), when necessary in the table which 
defines the reference configuration.

If the equilibrium configuration has symmetries, superposing the reference and bound configurations has not 
a single solution; shifting from a solution to an other by the ''symmetry operations'', is formally : i) a 
transformation of the coordinates; ii) a convenient permutation of the indices.

This classical explanation is not trivial, but it does not require postulates. As the transformations do not move 
the atoms, it is not necessary, when there are interactions with external fields, to move the whole universe 
or to correct, strangely, the symmetries.
\subsection{Computation of the energy levels}
When quantum mechanics started to develop, the quantum hamiltonians of simple systems were obtained by 
the correspondence principle, it led to partial derivative equations which were solved using standard 
computations. Later, it appeared that the resolution of the equations was easier using the raising and 
lowering operators; finally, it remained only Lie algebra (see an example of this evolution in \cite{Bobin}). 
Thus, one may think that the quantum theory is a method to introduce these algebra; however, the Lie 
algebra are now chosen arbitrarily, with the single aim to get good fits of the experimental results. The 
physical starting point is lost, in molecular, atomic spectroscopy, and elementary particle theories as well. 
It does not remain much of the old computations: a linear dependence of the energy of the one dimension 
harmonic oscillator; the remarkable function $j(j+1)$ for the energy of a rotator or a three dimensional 
harmonic oscillator comes from the isotropy of the space, the O(3) group and its algebra. Thus, the results 
which bring more than an interpolation over the discrete space of quantum numbers, come from the 
symmetry of the rotation (or, better, the Poincar\'e) group which must be taken into account in any problem 
of physics; the spin comes from the homomorphism of the SO(3) and SU(2) groups. More astonishing, the 
physicists use other interpolation methods where they work better, for instance the Pad\'e approximants in 
molecular spectroscopy of diatomic and ''spherical top'' molecules.
\medskip

Eigenvalues of the energy appear in most problems of classical mechanics because the potential energy has 
many relative minimums. Unhappily, their computations are often difficult, involving nonlinearities. We 
may use the phenomenological formalism of quantum mechanics as an interpolation method convenient to 
find these classical eigenvalues: the formalism of quantum mechanics gives approximate results of 
classical (or relativistic) mechanics.
\section{ Classical and quantum electrodynamics in the vacuum }
\subsection{The fundamental unconsciousness of quantum electrodynamics}
First, what is an optical mode for the linear Maxwell's equations? Set physically correct boundary conditions 
in particular that the fields have zero values for infinite space-time variables; then a mode is {\it any} 
solution of Maxwell's equations. Setting $w(\nu){\rm d}\nu$ the energy of the mode between the 
frequencies $\nu$ and $\nu + {\rm d}\nu$, the mode may be normalised by
\begin{equation}
\int_0^\infty \frac{w(\nu) {\rm d}\nu}{\nu}=h
\end{equation}
where $h$ is the Planck's constant.
Two modes are orthogonal if the energy of the system of the two modes is the sum of the energies of the two 
modes; a complete set of orthogonal modes (sentence often oversimplified into "set of modes") allows 
developing any mode on this infinite set of modes.

The fundamental postulate of quantum electrodynamics is the identification of a monochromatic optical mode 
with a harmonic oscillator. But, if the used set of modes is changed, the quantified energy of the modes is 
split \dots. A solution of this problem is ''the reduction of the wave packet", a postulate which allows the 
physicist to identify any mode with any other. This postulate is for me the strongest paradox, no, the 
strongest absurdity of quantum mechanics: how is it possible to work on objects whose definitions 
depend on the mood of the physicist?
\subsection{Some elementary, classical electrodynamics}
During a transition, a small, isolated mono- or polyatomic molecule emits a nearly monochromatic wave; an 
oscillating dipole (or quadrupole \dots) which radiates in nearly all directions generally models the 
radiating molecule. Suppose the frequency low enough to consider the source as small compared to the 
wavelength.

The electromagnetic energy in a sphere centred on the molecule, and whose radius is small compared to the 
wavelength allows evaluating the energy radiated by the molecule. If there is no external field, this energy 
is positive, the molecule loses energy; but an external field may cancel partly the molecular field, so that 
the molecule may absorb energy or radiate no energy; if it absorbs, it scatters too.

A consequence is that the energy $h\nu$ lost by a molecule cannot be absorbed by a single other molecule, 
the absorption requires an infinity of molecules, an infinite time: thus the universe is full of residues of 
electromagnetic fields, the ''stochastic'' or ''zero point'' field. It is fundamental to remark that this classical 
field is an ordinary electromagnetic field, that it must not be neglected, or studied independently of the 
other fields. The existence of the stochastic field explains, for instance, that the electron of the hydrogen 
atom generally does not lose energy although it radiates a field (except to reach the Lambshifted 
frequency).

The evaluation of the stochastic field requires the second Planck's law, not the first which, neglecting the 
stochastic field, sets that the energy in a mode is:
\begin{equation}
e=\frac{h\nu}{\exp(h\nu/kT)-1}.
\end{equation}
Supposing a high enough temperature, the exponential is developed:
\begin{equation}
e\approx \frac{h\nu}{h\nu/kT+(h\nu/kT)^2/2+…}\approx kT-h\nu/2
\end{equation}
From thermodynamics, it must be $kT$; thus $h\nu/2$ is added to the energy of the mode to get the second, 
good Planck's law. This energy is a stochastic electromagnetic energy in the mode \cite{Planck,Nernst}; 
its building shows that the corresponding field is an ordinary electromagnetic field; it provides the field 
which is necessary to compensate, in the average, the energy lost by radiation; if a system, such as a 
photoelectric cell requires a low energy to be excited, the long and particularly powerful fluctuations of the 
stochastic field are able to excite it, it is the noise observed at the lowest temperatures. Marshall and 
Santos \cite{Marshall} showed a local equivalence between the classical electrodynamics including the 
stochastic field that they call ''stochastic electrodynamics'', and quantum electrodynamics. Paradoxically, 
this equivalence is often used now to set that stochastic electrodynamics is an approximate fruit of 
quantum electrodynamics, while classical electrodynamics is older.

In the excitation or de-excitation of an electron, the electron provides the quantization of the electromagnetic 
field, just as the bottles quantify the wine. A common error is setting that this quantization is absolute 
while it applies only to systems that start from, and end at stationary states. The laws of refraction work at 
the lowest light levels while a lot of atoms are involved by the refraction by a prism. The temporary 
absorption of energy by each molecule of the prism during a light pulse is evidently much lower than 
$h\nu$. At the beginning of the pulse, when the field increases, the atoms get a slight excitation, remaining 
near their stationary state; if the prism is transparent, its atoms return their excess of energy to the tail of 
the pulse.

As the atoms are able to amplify the modes of the decreasing field, they are surely able to amplify other 
modes, but slightly (Rayleigh scattering) without the help of the coherence. Thus, the local stochastic field 
is increased during a pulse of light, the atoms and the fields get nearly an equilibrium, reversibly unless a 
big fluctuation of the amplified stochastic field excites an atom up to a transition.

In his thesis, Monnot \cite{Monnot} set a model of two energy levels atom, supposing that the amplitude of 
its radiating dipole is a quadratic function of the energy of the atom, equal to zero in the two stationary 
states. He puts a set of such atoms in a reflecting box; the energy of almost all atoms remains next to the 
eigenenergies, and the stochastic field keeps a nearly constant value; trying to increase or decrease the 
stochastic field is inefficient, provoking transitions of the convenient number of atoms. In conclusion 
''photon'' must mean "quantity of energy $h\nu$", not particle; W. E. Lamb \cite{WLamb} is not far from 
this point of view, but he does not make the last step, the rejection of the postulates of quantum 
electrodynamics.
\subsection{ Experiment of Einstein, Podolsky and Rosen in optics}
The emission of a photon is followed by an excitation of an atom by a photon only in the average. A source 
amplifies the stochastic field, so that the emission of a photon increases the probability that the 
fluctuations of the stochastic field grow up to values that pump atoms to higher states.

The paradoxes come from the quantization of the electromagnetic field; making a choice between the two 
locally equivalent theories, paradoxical quantum electrodynamics appears as an approximation of classical 
electrodynamics.

\subsection{Optical effects at low light levels}
Einstein considers two types of emission of light: the spontaneous emission, evaluated by the $A$ parameter 
and the stimulated emission evaluated by $B$. The experiments show that these parameters are not 
independent, the spontaneous emission being an emission stimulated by the stochastic field.

Suppose that a certain optical effect is a non-linear function $f(E)$ of the amplitude of the electric field of an 
incident beam of light; $E$ is produced by an amplification in a source of a stochastic field $E_0$ and may 
be written $E=E_0\beta$ where $\beta$ is the amplification coefficient of the source. It is $E$ which is 
written in $f(E)$ because the stochastic component of the field is an ordinary field which cannot be split 
from the remainder of the field.

If the light level is low, $\beta$ is nearly 1, so that
\begin{equation}
f(E)=f(E_0\beta)=f(E_0(1+\beta-1))\approx f(E_0)+(\beta-1)f ' (E_0).
\end{equation}
As the stochastic field is, in the average, constant, the effect is a linear function of the electric field, either 
$E=E_0\beta$, including the stochastic fraction, or $E_0(\beta-1)$ excluding it, as usual \cite{Moret4}. In 
particular, a photocell detects the available energy it receives, that is the difference between the received 
energy, including the stochastic field, and a restored stochastic field; at a low level, the signal is 
proportional to
\begin{equation}
E^2-E_0^2=(E_0\beta)^2-E_0^2\approx 2E_0^2(\beta-1).
\end{equation}
It is proportional to the amplitude, not to the intensity.

Seeming to ignore this elementary classical property, many authors gave wrong classical interpretations of 
experiments to show that quantum electrodynamics is ''the good electrodynamics''; they used photon 
counting to get sub-Poissonian statistics \cite{Short, Glauber}, or second order interferences:
\subsection{Second order interferences}
All proposed experiments (see, for instance \cite{Clauser,Gosh,Ou1,Ou2,Kiess}) are fundamentally equivalent, 
and they show only that their author's classical interpretations are wrong because the stochastic field is 
neglected.

Neglecting the stochastic field, the elementary explanation of these experiments \cite{M942} gives a contrast 
1/2 for the fringes, while the experimental value tends to one with a decreasing intensity of the light. 
Consider the simplest of these equivalent experiments \cite{Gosh}: two small photocells observe the 
interference of two small, incoherent, weak, monochromatic sources. These interferences are not visible 
because the sources are incoherent, their relative phase $\phi$ changes quickly, so that the fringes move 
too quickly for the eyes. The signal is the coincidences of the ''detected photons''.

Distinguish the two photocells by an index $j$ equal to 1 or 2, and the differences of the optical paths on the 
cell $j$ by $\delta_j$; a cell detects proportionnally to the amplitude $\cos(\pi\delta_j/\lambda+\phi/2)$, so 
that the probability of simultaneous detections is proportional to
\begin{equation}\cos (\frac{\pi\delta_1}{\lambda}+\frac{\phi}{2})\cos 
(\frac{\pi\delta_2}{\lambda}+\frac{\phi}{2}).
\end{equation}
Taking the mean value during an experiment, that is integrating $\phi$ on $2\pi$, we get a zero value for 
$\delta_1-\delta_2=\lambda/2$, so that the visibility reaches the right value 1.

For higher light intensities, the signal becomes proportional to the intensity, so that the visibility decreases to 
1/2.
\section{Tentative classical theory of the Wave particle duality.}
Quantum mechanics claim that it solves the problem of the wave particle duality; having the choice to 
consider an object as a particle or as a wave is not a true solution. The maser seemed absurd to people 
who considered the photon as a particle, but the next section will show that, in despite of the popularity of 
the lasers, the same arguments persist.

While the photon is not a particle, the electron, proton, neutron \dots are particles: they have a centre of mass, 
they may be static. The solitons are waves in non-linear media, which do not dissipate their energies by 
radiation; they are classified by a (p+q)D symbol, where p is the dimension of the wave, and q the 
dimension of its propagation. Unhappily our present mathematics allow  a rigorous study of a few (1+1)D 
solitons only. The (3+0)D solitons are fields which have the properties of particles: most of their energy is 
in a limited region of space, and this region may be static. Unhappily, these solitons do not seem to have 
been studied up to now. Show the existence of (3+0)D solitons.
\subsection{Known properties of optical solitons}
The nonlinearities which are generally studied in optics are produced by Kerr or photorefractive effects: the 
permittivity is an increasing function of the electric field, in many computations a quadratic function of this 
field up to a saturation provided by a sextic term.

The (1+1)D optical solitons have been extensively studied, they are probably used for the transmission of 
data in optical fibres.

The (3+1)D solitons, called optical bullets, propagate without dispersion, they are similar to particles which 
could not be stopped.

The optical (2+1)D solitons are light filaments obtained when a powerful laser is focalised in a non-linear 
medium \cite{Chiao,Marburger,Stegeman,Kivshar,Feit}.

The electromagnetic field decreases generally uniformly from the axis to the outside of a slightly converging 
laser beam, so that the index of refraction is larger near the centre of the wave surfaces; thus, the speed of 
light is lower near the centre, the curvature of the wave surfaces increases. If the beam is very neat, and 
the energy not too large, the beam converges to a single filament which is stable if it has exactly a ''critical 
flux of energy'', or radiates quickly an excessive energy. If the laser beam is powerful, local fluctuations 
provoke local convergences; many filaments are produced.

In a filament, the electromagnetic field may be artificially split into two parts: a cylindrical kernel in which the 
field is high, thus the nonlinearity large, and an evanescent wave whose amplitude decreases quickly with 
the distance to the kernel, so that it is quickly merged into external fields; the flux of energy in a stable 
filament free of interactions has the critical value, external fields may slightly change this flux. The 
properties of these solitons may be deduced from their theory or from their experimental observation.

We are interested here by the solitons in a perfectly transparent medium, but the absorption which occurs in 
real media brings useful informations: while the filaments lose much energy by molecular excitations, they 
are nearly parallel and so long \cite{Brodeur} that they surely absorb energy from the surrounding field to 
keep nearly the critical flux of energy; more precisely, there is an equilibrium between the external field and 
the filament: if the flux of energy in the filament is slightly under the critical value, it absorbs the 
surrounding field, and vice versa.

An other important result is deduced from the observation: the filaments do not merge, they often make 
regular figures, they repel each other; but, as they absorb their surrounding field, an interpretation of this 
repulsion is that the filaments are attracted to the regions where the field is the larger. The theory 
\cite{Zharov} and the experiments \cite{Petter,Shih} show that the filaments may be curved without a loss 
of stability by a non-uniform external field, either macroscopic, or created by an other filament.
\subsection{Theoretical existence of (3+0)D optical solitons}
We consider here perfect, isotropic, non-absorbing media having non-linear properties which enable the 
propagation of infinite filaments \cite{Moret3}. For a filament centred around $Oz$ with a given pulsation 
$\omega$, the electric and magnetic fields $\overrightarrow{E'}$ and $\overrightarrow{H'}$ are invariant 
by translations parallel to $Oz$, the lengths of which are integer products of a period $\Lambda$. Set that 
the evanescent field is negligible over a distance $\rho$ from $Oz$ 
\medskip

Consider another problem in which the medium has the previous properties and, in addition, a small, 
perturbing nonlinearity depending on the amplitude of the magnetic field. This perturbation does not 
destroy the stability of the filament.

Consider a circle $C$ of radius $R$ larger than $\rho$, whose circumference is an integer multiple of 
$\Lambda$. Set $\Omega$ a point of $C$, $\zeta_\Omega $ the curvilinear abscissa from $\Omega$ of a 
variable point $M$ of $C$, $M\xi$ an axis oriented to the centre of $C$ and $M\eta$ the axis making, with a 
tangent $M\zeta_M$ to $C$ a reference frame. Suppose that, for any point $M$, in a disk of radius $\rho$ 
and axis $M \zeta_M $, a daemon makes fields of amplitudes $E(\xi, \eta,  \zeta_\Omega )=E'(x, y, z)$ et 
$H(\xi, \eta,  \zeta_\Omega )=H'(x, y, z)$, oriented in $M\xi \eta  \zeta_M $ just as $ \overrightarrow{E'}$ 
and $ \overrightarrow{H'}$ in $Oxyz$.
\medskip

As the pulsation $\omega$ is a constant, the perturbation may be considered as a function of the amplitude of 
the curl of the electric field, rather than a function of the amplitude of the magnetic field.

Set $\Pi_M$ the plane orthogonal to $C$ at abscissa $ \zeta_\Omega $ and $\Pi_N$ a similar plane at a point 
$N$ of slightly larger abscissa $ \zeta_\Omega +\delta  \zeta_\Omega $. Set $\alpha$ the angle of rotation 
of the tangent to $C$ from $ M$ to $N$, that is the angle between $\Pi_M$ and $\Pi_N$.

In a second order approximation, the component along $M\eta$ of the curl of $ \overrightarrow{E}$ supposed 
polarised along $M\xi$, is $\delta E_\xi(\xi, \eta,  \zeta_\Omega , t)/( \delta  \zeta_M)=\delta E_\xi(\xi, \eta,  
\zeta_\Omega , t)/( \delta  \zeta_\Omega (1+\xi\alpha))\approx(1-\xi\alpha) \delta E_{\xi}(\xi, \eta,  
\zeta_\Omega ,t )/ \delta  \zeta_\Omega $. As $\delta  \zeta_M$ is a decreasing function of $\alpha \xi$, the 
amplitude of the curl and the index of refraction increase with $\alpha \xi$.

Huyghens' construction applied to the wave in plane $\Pi_M$ leads to a distorted and, in the average turned 
wave surface. Happily Huyghens' construction is too imprecise in a filament whose stability shows that 
the wave surface is not distorted; however, as the variation of the the index of refraction is odd in $\xi$, the 
wave surface is turned by an angle $\beta$.

The function $f(\alpha)=\beta(\alpha)/\alpha$ may be adjusted, using the variation of the index of refraction as 
a function of $ \overrightarrow{H}$, so that, for a certain value $\alpha_0$ of $\alpha$, $f(\alpha_0)=1$, 
${\rm d}f(\alpha_0)/ {\rm d}\alpha<0$. The daemon is not anymore useful, an autocoherent and stable 
solution is found. The filament is transformed into a torus that traps the electromagnetic field; as the 
length of the filament and the flux of energy are fixed by the optical parameters, the energy of the soliton is 
quantified.
\medskip

The existence of (3+0)D solitons is demonstrated, but a true study seems to require very powerful computers, 
with, maybe, the previous torus as starting point. If the properties of the filaments remain true, two toruses 
repulse each other; the regions where the field at the same frequency is large attract the torus.

Some crystals, tourmaline for instance, have remarkable electric and magnetic properties, but they absorb the 
light so much that it seems impossible to use them to try optical (3+0)D solitons. Are the balls of fire 
produced by the lightnings optical solitons? They seem made of ionised gas that could have the required 
properties.
\subsection{Are particles optical solitons ?}
Purely electromagnetic interactions, in the $\gamma$ range can produce electron-positron pairs \cite{Ritus}; 
this shows that the vacuum becomes nonlinear. Remark that quantum mechanics introduce such 
nonlinearities through virtual particles. Testing whether matter is made of electromagnetic solitons is a big 
job! Suppose it is true.

If the kernel of a soliton goes through a Young hole, its evanescent wave propagates through both holes. 
Over the screen, we have a superposition of incoherent fields and of the interferences coming from the 
evanescent field; the incoherent fields do not interact much with the soliton; the soliton moves to the 
regions where the interferences are bright. Thus, the torus may be de Broglie's $u$ field while the 
evanescent field is the $\psi$ field \cite{deBroglie}.

A lot of different (3+0)D solitons may be defined in a single medium, using, for instance, the following 
methods:

- Changing the frequency (consequently $R$) and the polarisation of the wave;

- Commuting the roles plaid by the electric and magnetic fields;

- Using as index of refraction a function of the amplitudes of the fields which has many maximums;

- Introducing a torsion of the curved filament.
\section{Quantum mechanics: a source of errors in astrophysics.}
Quantum mechanics pretends find a solution to the wave-particle duality, but it is difficult to understand its 
rules: Why do a single atom absorbs a photon, while all atoms of a prism are necessary to explain the 
refraction of a single photon? Saying that the maser cannot work, the best physicists are not conscious to 
make arbitrarily the choice ''particle''. For this problem, the other choice works, but the problem of the 
locality appears \dots.

	Unhappily, the astrophysicists made a wrong choice: Forbidding by a rejection of the coherent effects the 
alternatives to the Doppler (or expansion) effect they are obliged to imagine extraordinary interpretations 
of observations.
\subsection{Coherent Raman scatterings.}
''Coherent Raman scattering'' is usually used for the scattering of laser pulses, with frequency shifts. In the 
region reached first by the laser beam, the emission is spontaneous, then the scattered light is amplified. 
The process is nonlinear so that few Raman line appear. The experiments show coherence between the 
incident light and the ability of matter to amplify a Raman frequency. To maintain the coherence of phase 
in despite of the dispersion between the incident and scattered lights, their directions of propagation must 
generally be different, a cone of scattered light is obtained.

This possibility fails if the beam is wide, because the wave surfaces are identical for the incident and scattered 
lights: the amplification is limited to a ''coherence length''. At the limit, for Rayleigh scattering, the 
coherence length is infinite and the refraction produced by an interference of the incident beam with the 
($\pi/2$ phase shifted) coherently scattered beam, is a large effect.

The ''Impulsive Stimulated Raman Scattering'' (ISRS), known since 1968 \cite{Yan} is used mostly in chemical 
physics \cite{Nelson,Weiner,Dougherty}; it uses ultrashort laser pulses. ''Ultrashort'' has two meanings: it 
may be relative to the shortest available pulses (usually femtosecond pulses) or to the physics of their 
interaction with matter: following G. L. Lamb \cite{Lamb} ''shorter than all relevant time constants''.

\medskip
The name ''Impulsive Stimulated Raman Scattering'' is not very convenient because ''Raman effect'' 
corresponds to a two photon effect in which the initial and final levels differ. ISRS is a four photons effect 
which does not pump the molecules. However this parametric effect may be considered as a combination 
of two, nearly simultaneous Raman effects.

With Lamb's definition of the ultrashort pulses, the time between the collisions must be longer than the length 
of the pulses, so that the coherence of the scattering is not perturbed by the collisions even if the pulses 
are weak. The period of the beats between the incident and scattered light is larger than the length of the 
pulses, so that the beams interfere into a single, monochromatic, but frequency-shifted beam; this type of 
interferences is usually observed with two lasers, or in a Michelson interferometer, when a mirror moves. 
Thus, a frequency shift is obtained without any blur of the spectral line or of the images. However, as the 
power is large, the scattered amplitude is generally proportional to the square of the incident amplitude 
(stimulated scattering), so that the relative frequency shift is proportional to the intensity: the lines of a 
polychromatic spectrum have different relative frequency shifts.
\subsection{Incoherent Light Coherent Raman Scattering (ILCRS)}
Natural, incoherent light is made of pulses, the length of which is of the order of 10 nanoseconds.
\medskip

The first condition to consider it as made of ultrashort pulses is that the collisional time be longer enough 
than 10 ns; it is an ordinary vacuum, so that the observation of an effect requires a long path.
\medskip

The second condition is that the period of the beats of the incident and scattered lights are longer enough 
than 10 ns: The Raman transition must be in the microwaves range, that is the active molecules must have 
hyperfine structures.
\medskip

As the power of natural sources is low, the amplitude scattered by ILCRS is proportional to the incident 
amplitude, the relative frequency shift depends on the frequency by dispersion effects only. The intensity 
of a parametric effect made of the combination of two effects, is limited by the weaker effect; here the low 
energy Raman transition, so that, in the visible, the relative frequency shift is almost constant.

\medskip
ILCRS does not blur either the spectra or the images; it introduces a very nearly constant relative frequency 
shift: from the spectra and the sharpness of the images, {\it it is very difficult to distinguish ILCRS 
redshifts from Doppler redshifts}.
\subsection{Application to astrophysics}
It seems difficult to observe ILCRS in the labs, but astrophysics provides low pressures and long paths.

Which molecules may be this sort of catalyst, able to transfer energy from the redshifted high frequencies to 
the slightly heated thermal radiation at 2.7K?

All mono or polyatomic molecules are able to acquire Stark or Zeeman hyperfine structures; often, heavy 
molecules have such structures, but their density is low. The nuclear spin coupling transition of H$_2$ at 
.21 m is weak, but the molecules which have an odd number of electrons have strong transitions in 
hyperfine structures. NO, OH, NH$_2 \dots$ have been observed in the galaxies.

The ultraviolet radiations can ionise H$_2$ into H$_2^+$, a very stable molecule where the collisions which 
destroy it are rare. It is easy to understand that the observation of this molecule is difficult: it has many 
relatively weak lines which are spread by the redshift produced by the molecules themselves; increasing 
the pressure, H$_2^+$ is destroyed by collisions before a decrease of the ILCRS redshift.
\medskip

The two main proofs of the expansion of the universe, the cosmological redshift, and the existence of the 2,7K 
radiation should be tested against their production by an ILCRS effect. A number of H$_2^+$ molecules, 
of the order of 20 in a cubic metre would produce the whole cosmological redshift \cite{Moret,Moret1}.

The small dispersion of ILCRS could explain the dispersions observed in the spectra of quasars \cite{Moret5} 
without the hypothesis of a variation of the constant of fine structure \cite{Webb}.

Near some bright QSOs, powerful thermal radiation observed in the infrared seems produced by heated dust, 
but the pressure of radiation should push the dust out. Is it an ILCRS effect?

A lot of theories tried to explain the multiplicity of the Lyman lines in the spectra of many quasars; ILCRS 
proposes a purely static, simple explanation supposing a variable magnetic field in a halo: where the field is 
low, the lines are written into the spectrum; elsewhere, ILCRS is activated by the Zeeman effect, the 
spectral lines are so spread by the shift that they are invisible \cite{Moret2}.
\section{Conclusion}
Wanting to understand anything, we are tented to use marvellous explanations; the science needs centuries 
to destroys the marvellous by theories which are coherent and verified by the experiments. 

The usefulness of the quantum theory is not a proof of the rightness of its postulates that the paradoxes 
show fundamentally incoherent; attributing the formalism of quantum theory to the classical theory seems 
possible, with the strong advantages of precision and lack of paradoxes.

Unavoidably wrong interpretations of quantum theory led the astrophysicists to neglect the search of an 
optical theory to interpret a part of the redshifts, thus introducing strange hypothesis, unable however to 
explain hundreds \cite{Reboul}, probably now more than thousand observations.

\end{document}